\documentstyle[11pt,paspconf,epsf]{article}
\def\kpch{\,{h^{-1}{\rm Kpc}}}

\def\kms{\,{\rm {km\, s^{-1}}}}

\def\vcir{{V_{\rm c}}}
\def\nhi{{N_{\rm HI}}}
\def\nmgii{{N_{\rm MgII}}}
\def\nciv{{N_{\rm CIV}}}
\def\novi{{N_{\rm OVI}}}
\def\cm{\,{\rm{ cm}}}
\def\kpch{\,h^{-1}{\rm {Kpc}}}
\def\zsun{Z_{\odot}}
\def\J{J_{-21}}

\def\edcomment#1{\iffalse\marginpar{\raggedright\sl#1\/}\else\relax\fi}
\reversemarginpar

\begin{document}
\title{Gaseous Galactic Halos and QSO Absorption Line Systems}
\author{H.J. Mo}

\affil{Max-Planck-Institut f\"ur Astrophysik, 85748 Garching, Germany}

\begin{abstract}
  Recent observations have shown that (some) metal line absorption systems 
in QSO spectra arise from gas in galactic halos with radii
much larger than the optical radii of galaxies. We show that
the observed galaxy-absorber connections are natural results 
of galaxy formation in hierarchical cosmogonies, where galaxies
form from gas cooling and condensation in dark matter halos.
\end{abstract}

\section{Introduction}

  There is now much evidence that metal line absorption systems in QSO
spectra arise from gas in galactic halos.
Recent imaging and spectroscopic observations can
identify directly, at low redshift ($z\la 1$), the galaxies associated with the
MgII absorption systems (e.g. Bergeron, Cristiani, \& Shaver 1992; 
Bechtold \& Ellingson 1992; Steidel 1995; see Steidel 1998 for a 
recent review). The properties of such galaxy/absorber systems 
can be briefly summarized as follows:
(i) Most MgII systems with rest-frame equivalent widths $W>0.3$\AA~
are associated with galaxies;
(ii) Most galaxies with ($K$-band) luminsity $L_K\ga 0.05 L_K^*$ 
       have gaseous halos to produce MgII systems with $W>0.3$\AA; 
(iii) The typical absorption radius of a galaxy with $L_K\ga 0.05 
       L_K^*$ is $R_a\sim 35\kpch(L_K/L_K^*)^{0.15}$, 
       much larger than the optical radius;
(iv) Galaxies with $L_K\la 0.05 L_K^*$ have much
       smaller absorption cross sections than the brighter ones;
(v)  For bright galaxies, the absorbing clouds has   
       roughly spherical distribution, with a covering factor 
       $\ga 1$ within $R_a$, as indicated by the fact that
       few interlopers are observed;
(vi) The absorbing galaxies show wide range of colors, from 
      that of late type spirals to that of early type ellipticals, 
      indicating that bright galaxies of all types
      possess similar gaseous halos;
(vii) The observed absorption line systems show velocity 
      structures, with typical velocity spread of $\sim 
      50$-$200\kms$ (e.g. Petitjean \& Bergeron 1990; 
      Churchill 1998), consistent with the absorption clouds
      moving in the potential wells of galaxies;
(viii) The absorbing gas is enriched, as heavy elements are observed. 

  Despite of the many observational facts about the relations
between galaxies and the metal line absorption systems, 
relatively few theoretical work has been done. 
In particular, the importance of such observational results to 
our understanding of galaxy formation has not been fully explored. 

  In hierarchical clustering models of galaxy formation, it is 
assumed that galaxies form at the centers of dark matter halos as 
gas collapses, is shocked, and cools. 
Although our knowledge of how galaxies form is still
incomplete at present time, what is certain is that at some
time in the past gas had to dissipate and move from the outer parts 
to the inner parts of the dark matter halos around galaxies, 
and in this process it must have produced absorption line systems in 
any quasar line-of-sight intercepting galactic halos.
It is clear that in order to understand galaxy formation we must first
understand the physical state of the gas out of which galaxies form,
and that absorption line systems are an excellent observational probe
to any gas dissipating in halos. Thus, the observational evidence 
that the high column density absorption
systems at $z\la 1$ indeed arise in gaseous halos around galaxies
should be a landmark for our understanding of galaxy formation.
  Based on these considerations, we have proposed 
a simple two-phase model for the gaseous structure of 
galactic halos (see Mo, 1994; Mo \& Miralda-Escud\'e 1996).
The model is based on current scenario of galaxy 
formation in hierarchical clustering models, and so 
allows us to understand the absorption line systems 
in the general framework of structure formation in the 
universe. It was found that the observed galaxy-absorber connections 
are natural results of galaxy formation in hierarchical 
cosmogonies.

\section{The Structure of Gaseous Galactic Halos}

During the process of galaxy formation, intergalactic gas
collapse and move inward through the extended dark matter halos that
are observed around present galaxies. A halo of hot gas at the virial
temperature will form as the kinetic energy of the infalling material
is thermalized in shocks. In low mass halos, the cooling time of this
hot halo gas is short compared to the dynamical time if it contains
all the accreted baryons; therefore, the gas must cool until the density
of hot gas decreases to a level such that the cooling time is similar
to the age of the system. Due to the increased cooling rate as
temperatures drop, a two-phase medium will naturally form in these 
conditions of rapid cooling. Clouds photoionized by a radiation
background will be maintained at an equilibrium
temperature of about $10^4$ Kelvin, in pressure equilibrium with 
the hotter halo gas. These clouds could form from 
inhomogeneities in the halo gas, or from the ram-pressure 
stripped interstellar medium of satellite galaxies. Due to their 
higher density, clouds will fall to the center of the halo where they
may form globular clusters or settle into gaseous disks. 
In such a scenario, modeling the structure of gaseous galactic
halos involves both the formation of dark matter halos and
the state of gas in dark matter halos. The following is a sketch
of the main ingredients involved in such a modeling.

\subsection {Dark halos}

  The properties of dark matter halos in hierarchical clustering
cosmogonies [such as cold dark matter (CDM) models] are relatively 
well understood from analytic models and N-body simulations.  
 Dark matter halos can be described roughly as singular isothermal 
spheres, with density profiles given by 
\begin{equation}
\rho(r) = \frac{\vcir^2}{4\pi G r^2};\,\,\,\,\,\,\,\,\,
\vcir=\left(\frac{G M}{r_v}\right)^{1/2}.
\end{equation}
where $\vcir$ is the circular velocity, $M$ is the
mass, and $r_v$ is the virial radius, of the halo.
From spherical collapse model, the virial radius can be defined as
$r_v={\vcir}/{10 H(z)}$, 
where $H(z)$ is the Hubble constant at redshift $z$.  

 For a given cosmogonic model, one can also estimate the 
mass function of dark matter halos, $n(M;z) dM$, which gives
the comoving number density of dark matter halos, 
with mass in the range $M\to M+ d M$, at a given redshift $z$.
This function can be estimated either from N-body simulations,
or from analytic models such as the Press-Schechter formalism. 
Thus, the number of halos which may host
absorption line systems can be obtained.
  
 As demonstration, results in the following are shown 
only for the standard CDM
cosmogony with $\Omega_0=1$, $h=0.5$, and $\sigma_8=0.67$.

\subsection {Gas in Dark Matter Halos}

Before the collapse of a dark matter halo, gas on a mass shell
moves together with dark matter particles until the velocity 
of the mass shell reaches the sound speed of the gas interior
to the mass shell, where the gas is shocked while dark matter 
particles continue to collapse and virialize. Ananlytic models
(e.g. Bertchinger 1989) and numerical simulations 
(e.g. Evrard 1990) show that, if cooling is not efficient
then the gas in a dark halo will be shock-heated to the virial 
temperature
\begin{equation}
T_v=\mu {\vcir^2\over 2k}, 
\end{equation}
with a profile similar to that of the dark halo. 
 Because the gas density is higher near the center of a
halo, radiative cooling of the gas is more efficient there. We 
define a cooling radius $r_c$ by 
\begin{equation}          
f\rho(r_c)={5\mu k T_v \over 2\Lambda (T_v)t_M},
\end{equation}
where $\Lambda(T)$ is the cooling rate, $\mu$ is the average mass
per particle, $f$ is the gas mass fraction and $t_M\sim t/2$
is the time between major mergers. 
Thus, gas located at $r\ll r_c$ can cool effectively
before the halo merges into a larger system, 
while that at $r\gg r_c$ cannot cool and
should retain its original state. Based on this consideration,
we model the density profile of the hot component as
\begin{equation}
\rho_{\rm hot}(r)=\frac{f\vcir^2}{4\pi G r(r+r_c)},
\end{equation}
so that $\rho_{\rm hot}(r)\to f\, \rho(r)$ for $r\gg r_c$
and $\rho_{\rm hot}(r)\propto 1/r$ for $r\ll r_c$. 
The temperature of this phase is assumed to be at 
$T_{\rm hot}\sim T_v$. These behaviours of $\rho_{\rm hot}(r)$ are 
similar to those obtained from the self-similarity solution given by 
Bertchinger (1989). When $r_c>r_v$, the cooling time at the virial 
radius is already short compared to $t_M$. In this case we 
assume $r_c=r_v$.

   The accreted gas that does not stay in the hot phase and cools
must fall through the hot halo in the form of photoionized clouds.
We refer to the gas in these clouds as the ``cold phase''.
The rate at which cold gas accumulates in a halo is determined by 
both gas infall and gas cooling, and is roughly
\begin{equation}
{\dot M}_{\rm cold}\sim \frac{f\vcir^2}{G t_{\rm M}}
r_{\rm min};\,\,\,\,\,\,\,\,  r_{\rm min}={\rm min}[r_v, r_c].   
\end{equation}
We assume the mass flow rate to be ${\dot {M}}
=M/t_M$, and that the clouds move to the halo center with a constant
velocity $v$.
Assuming also spherical symmetry for the gas distribution,
we can write the density of the cold gas as a function 
of the distance $r$ to the halo center as
\begin{equation}
\rho_{\rm cold} (r) = {\dot M \over 4\pi r^2 v }.
\end{equation}
In the absence of a hot phase, the cold gas is in free fall and $v$
must be of the order of the virial velocity $\vcir$. The friction of hot
gas may cause cold clouds to move at a terminal velocity which is
smaller than $\vcir$. However, since halos on galactic scales have
cooling times that are similar to the dynamical times, the inflow
velocity should not be much smaller than the virial velocity.
This is also consistent with the observed velocity structure
in MgII systems.  
The cold clouds are assumed to be at a temperature
$T_{\rm cloud}\sim 10^4$K, they are also assumed to be
in pressure equilibrium with the hot gas, so that
the density of a cloud (assumed to be uniform) at a radius 
$r$ is
\begin{equation}
\rho_{\rm cloud}\sim 
\rho_{\rm hot}(r)T_{\rm hot}/T_{\rm cloud}.
\end{equation}

\subsection {Cloud properties}

  In the model outlined above, cold clouds are moving in 
diffuse substrates of hot gas. Various physical processes 
can act to affect the formation and destruction of the clouds.
These processes constrain the masses and densities of the 
clouds. When the mass of a cloud is too large and its density 
too high, the cloud will become gravitationally unstable as it
sinks towards the halo center. This happens when the cloud mass 
exceeds the Jeans mass. Once a cloud is accelerated to a terminal 
velocity at which the drag force of the hot medium is comparable
to the self-gravitating force on its surface, the cloud may become
hydrodynamically unstable. Small clouds are also unstable against
heat conduction. Based on these considerations, we found that 
the masses of cold clouds are confined to a narrow range,
$M_{\rm cloud}\sim 10^6 M_{\odot}$. The typical size 
of such a cloud is $\sim 1\,{\rm K pc}$. The covering 
factor of such clouds is $\sim 1$ within $\sim 30\kpch$ 
for halos with $\vcir\sim 200\kms$. 

\section {Implication for QSO Absorption Line Systems} 

  To make connections between gaseous halos and QSO absorption
line systems, we need to calculate the column densities of
ions (of some species) at different impact parameters from a halo. 
We assume that the gas is ionized by the UV background 
radiation with a flux $J(\nu)$ which we parameterize as 
\begin{equation}
J(\nu)=\J (z) \times
10^{-21}\left( {\nu \over \nu_{\rm HI} }\right)^{-\alpha} \Theta (\nu) \,
{\rm erg\,cm^{-2} sr^{-1} hz^{-1} sec^{-1}},
\end{equation}
where $\nu_{\rm HI}$ is the hydrogen Lyman limit frequency, $\J (z)$ gives
the redshift dependence of $J(\nu)$, and $\Theta(\nu)$ describes
departures of the spectrum from a power-law. 
A typical spectrum  can be represented approximately by taking
$\alpha = 0.5$, $\J (z)=0.5$ for $z>2$ and
$\J (z) = 0.5\times [(1+z)/3]^{2}$ for $z<2$,
and including a break in the spectrum at $\nu _4\equiv 4$ Ryd (due to
continuum absorption by HeII), with $\Theta(\nu<\nu_4)=1$
and $\Theta_4\equiv \Theta(\nu\ge \nu_4)=0.1$.

To consider metal line systems, we also need to assume a metallicity, 
$Z$, of the gas. The exact level of chemical enrichment in 
galactic halos is not clear. In the disk-spheroid model of our
Galaxy, the metal output from stars in the spheroid component
can enrich the disk gas to a level of about $0.2\zsun$ (see
Binney \& Tremaine 1987, \S9.2). In this model, the gas in the halo 
was initially enriched to such a level before it settles into the disk. 
In the galaxy formation model of White \& Frenk (1991), gas in halos
with $\vcir\ga 100\kms$ could also be enriched to a level of 
$0.1$-$0.3\zsun$ (see their Fig. 4), because of star formation
in the progenitors and that in halos themselves. Unfortunately,
the detail result depends crucially on how well the output metal is 
mixed with halo gas. In our model we treat $Z$ as a free parameter.
 
With this assumptions, the number density of any species of ions 
can be obtained as a function of the radius from the center of
a halo with circular velocity $\vcir$ at redshift $z$. 
As a result, for single halos we can obtain 
$N_X(D)$, the column density of species $X$ expected at 
an impact parameter $D$, and $D(N_X)$, the impact parameter 
at which the column density is $N_X$. Given the number density
of dark halos as a function of $\vcir$ and $z$ 
(e.g. from the Press-Schechter formalism),
we can also obtain the statistical properties of 
absorption systems.

 The left panel of Figure 1 shows the HI column density
$\nhi$ versus impact parameters for halos at $z=0.5$ with various 
$\vcir$. The value of $z$ chosen here is to match the mean 
redshift of the MgII systems for which associated galaxies 
are identified (e.g. Steidel 1995). As one can see,
$\nhi $ decrease with $D$. 
This is because $\nhi$ is proportional to the total column 
density of cold gas, which is higher for a smaller
impact parameter, and to the pressure (since neutral fraction 
increases with pressure), which is higher near the center. 
The figure shows that the impact parameter for producing a Lyman
limit system (LLS) with $\nhi\ga 10^{17} \cm^{-2}$ is 
about $40\kpch$ at $\vcir\sim 200\kms$. 
In halos with $\vcir\ga 250\kms$, this impact parameter is not much 
larger because the cooling radius is reached, and we assume that no 
clouds are formed outside $r_c$.

\begin{figure}
\plottwo{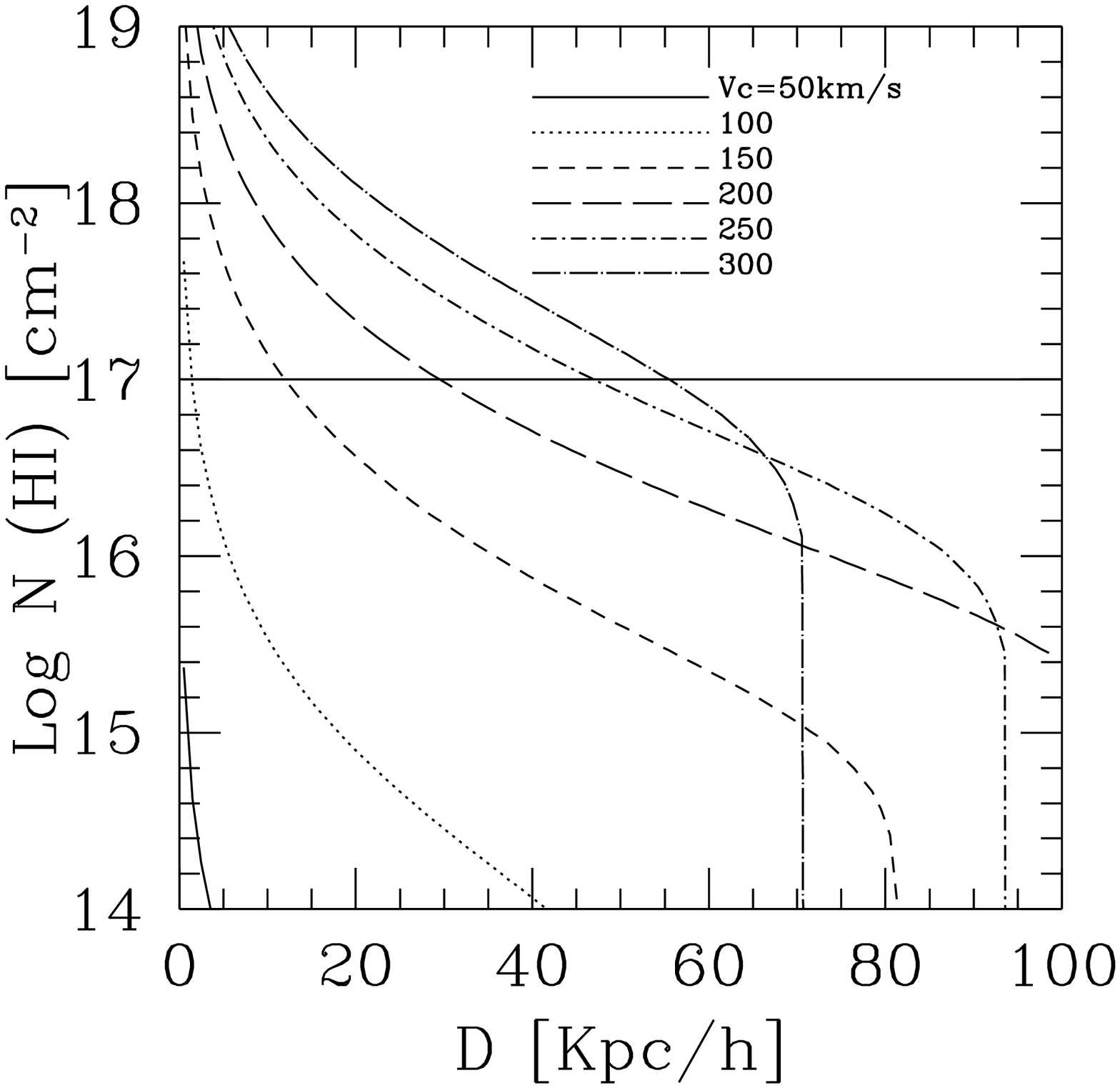}{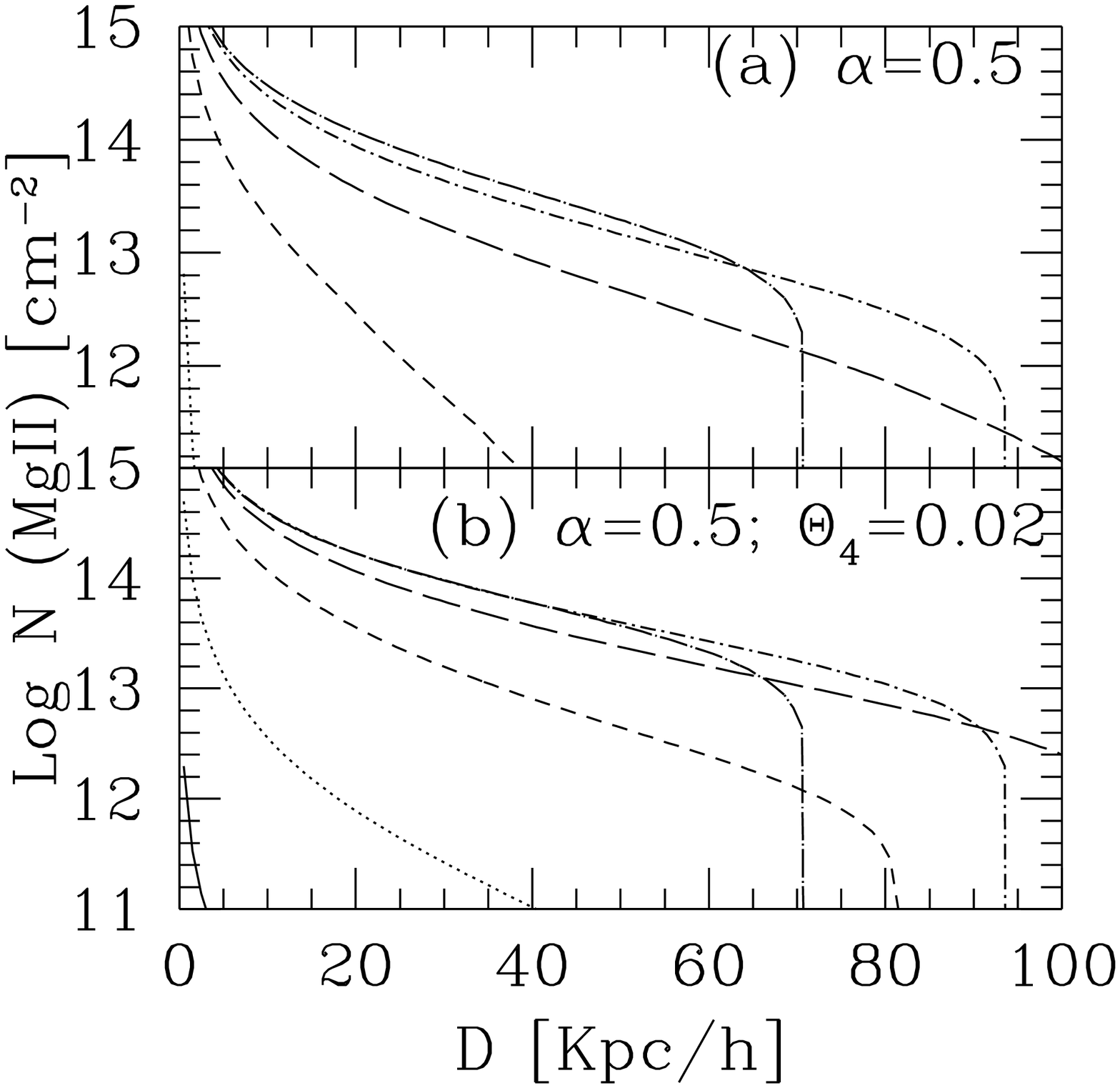}
\caption {{\bf Left panel}: HI column density as a function of impact 
parameter $D$, for halos at $z=0.5$ and with various $\vcir$. 
The model parameters are $f=0.05$, $Z=0.3\zsun$ and $v=\vcir$. 
The horizontal line indicates $\nhi=10^{17}\cm ^{-2}$, above 
which an LLS is produced.
{\bf Right panel}: MgII column density as a function $D$.
Results are shown for two models of UV flux, one is a power law
with power index $\alpha=0.5$, the other has the same power index but 
with a break at 4 Ryd so that $\Theta_4\equiv \Theta (\nu\ge \nu_4)=0.02$.
In both cases $J_{-21}=0.1$.} 
\end{figure}

\begin{figure}
\plottwo{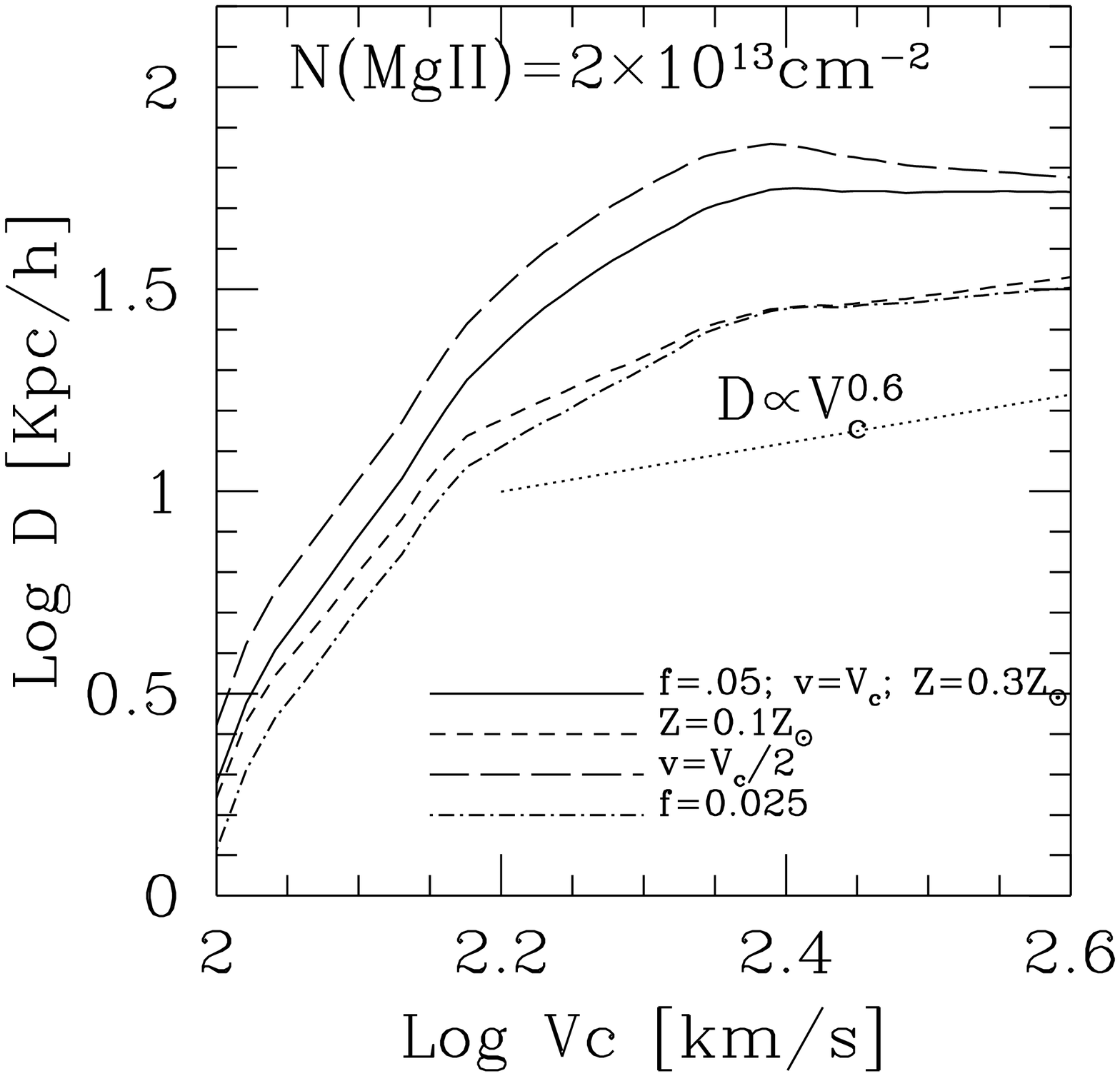}{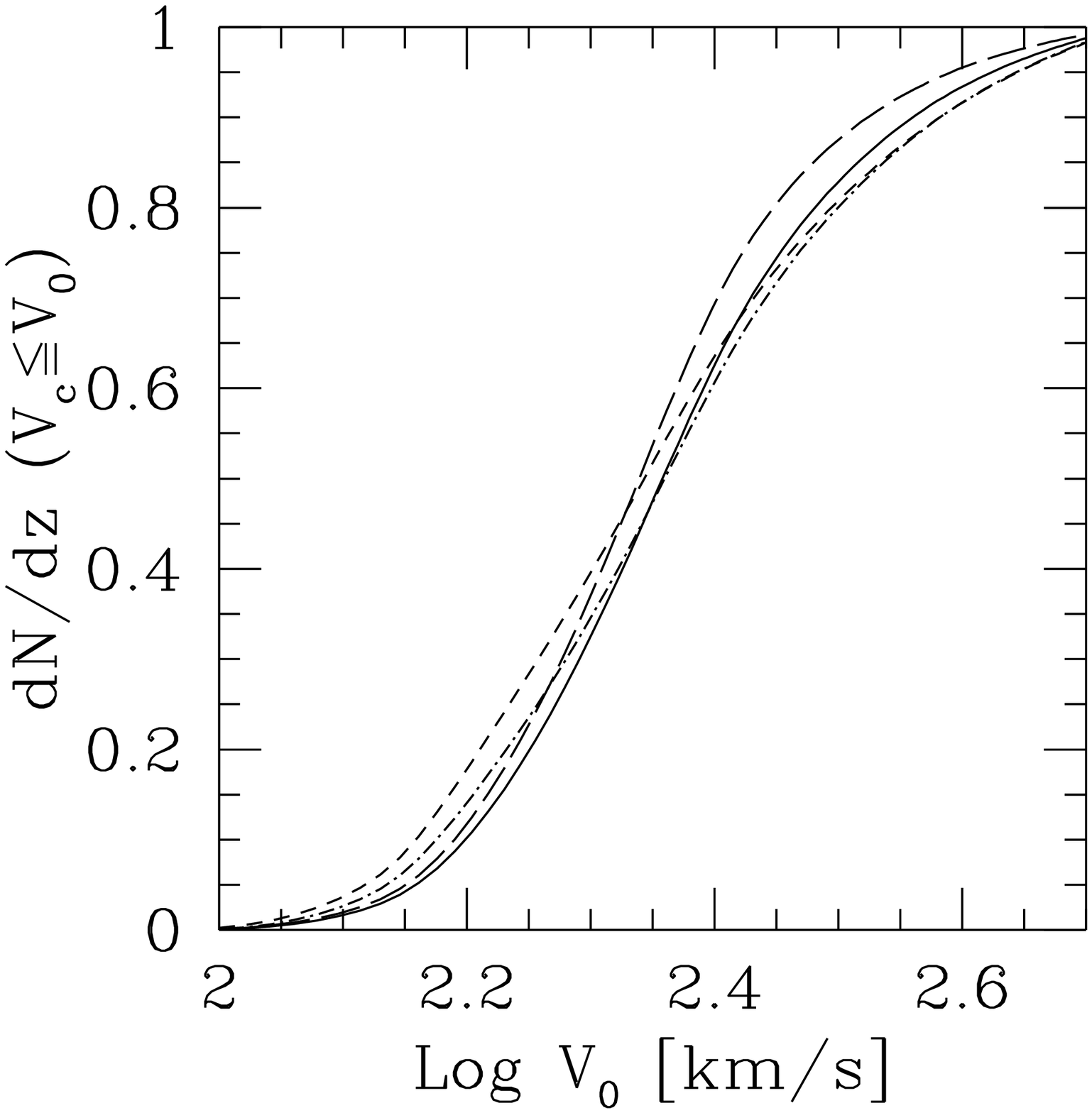}
\caption {{\bf Left panel}: 
The impact parameter $D$, at which $\nmgii=2\times 10^{13}\cm ^{-2}$,
as a function of halo circular velocity. 
Results are show for halos at $z=0.5$.
The dotted line shows the relation $D\propto \vcir^{0.6}$. 
{\bf Right panel}: 
Cumulative cross section of MgII systems at $z=0.5$, with
$\nmgii\ge 2\times 10^{13}\cm ^{-2}$, given by halos with circular
velocity $\vcir\ge V_0$, as a function of $V_0$.}
\end{figure}

\begin{figure}
\plottwo{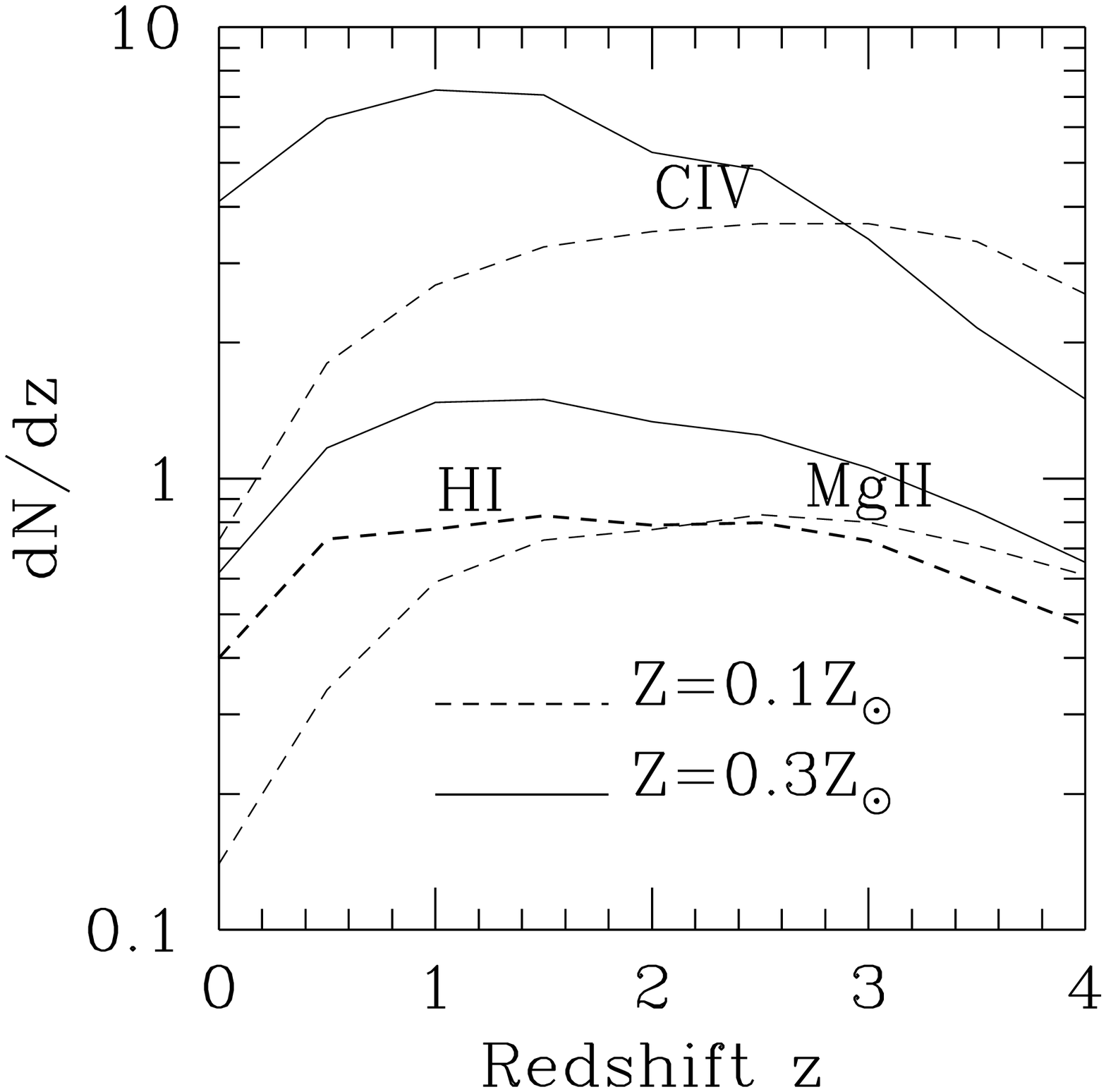}{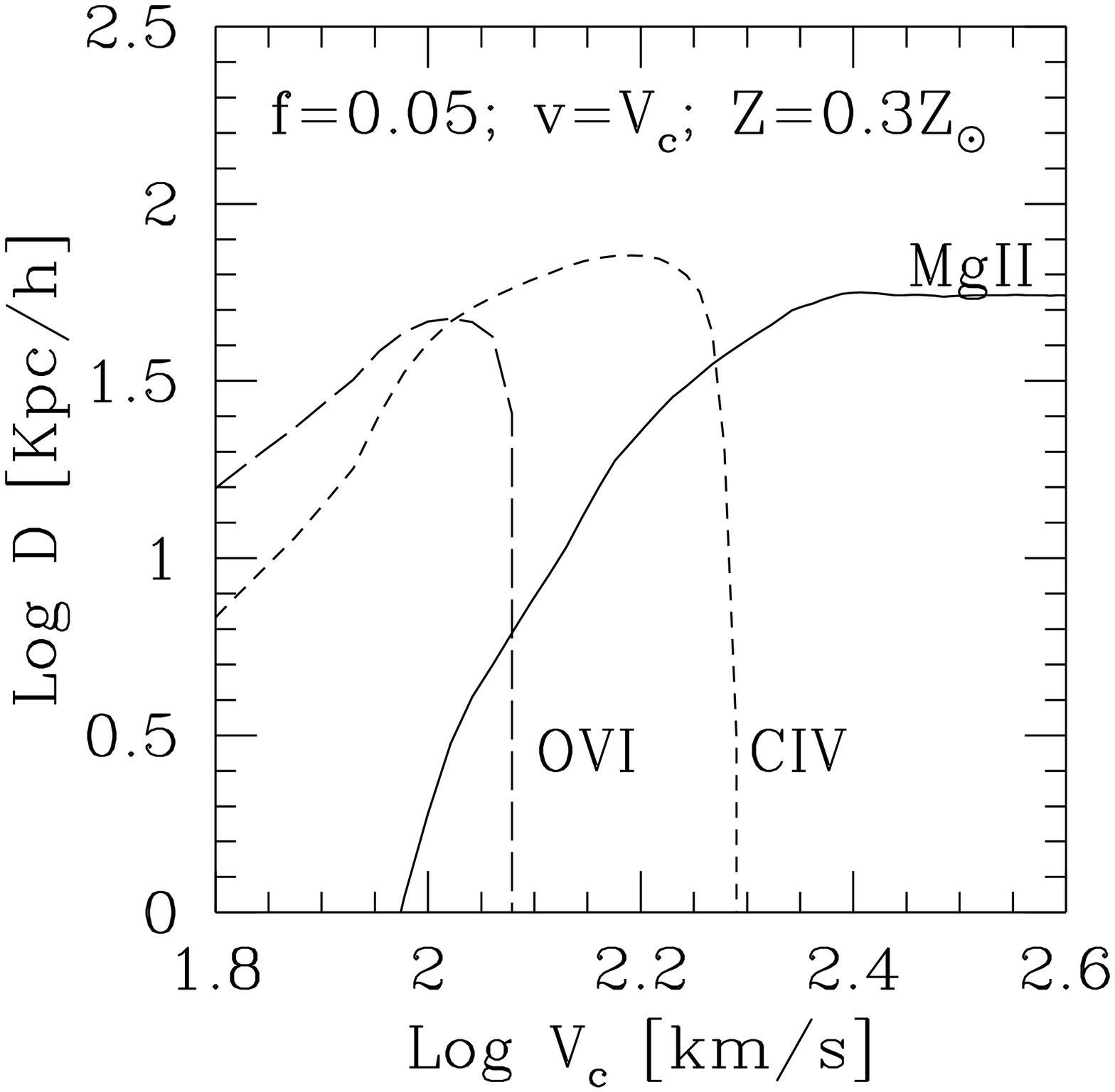}
\caption {{\bf Left panel}: Number of absorption systems per unit redshift,
for HI systems with $N_{\rm HI}\ge 10^{17}{\rm cm}^{-2}$, MgII
systems with $N_{\rm MgII}\ge 2\times 10^{13}{\rm cm}^{-2}$, and
CIV systems with $N_{\rm CIV}\ge 10^{14}{\rm cm}^{-2}$.
Solid curves assume $f=0.05$ and $v=\vcir$ with $Z=0.3\zsun$;
dashed curves assume the same $f$ and $v$ but with $Z=0.1\zsun$. 
{\bf Right panel}: 
The impact parameters at which 
$\nmgii=2\times 10^{13}\cm ^{-2}$ (solid),
$\nciv= 10^{14}\cm^{-2}$ (short dashed), 
and $\novi =10^{14} \cm ^{-2}$ (long dashed), as a function of 
$\vcir$ for halos at $z=0.5$.}
\end{figure}

The shape of the MgII column density profile 
(shown in the right panel of Fig.1) is similar to that of HI,  
because $\nmgii/\nhi$ is roughly a constant  
for the photoionization model considered here.
The observed samples of MgII absorption lines select systems where the
MgII line is above a threshold of equivalent width. 
We present results for the number of absorbers
and the impact parameter distribution assuming a fixed threshold
$\nmgii = 2\times 10^{13}\cm ^{-2}$ (reasonable for $W = 0.3$\AA), 
although we should bear in mind that the average cloud covering factor
can also affect the observed equivalent widths.
The left panel of Figure 2 shows the results for the impact 
parameter in halos of different $\vcir$, at redshift $z=0.5$.
The UV flux spectrum used here has parameters 
$J_{-21}=0.1$, $\alpha=0.5$ and $\Theta (\nu\ge\nu_4) =0.1$. 
At $\vcir\la 150\kms$, $D$ decreases rapidly with decreasing $\vcir$, 
due to the faster cooling rate and the lower pressures that are 
implied. The typical impact parameter is 
about $30\kpch$ at $\vcir\sim 200\kms$.
In halos with $\vcir\ga 250\kms$, this impact parameter is not much 
larger again because the cooling radius is reached and
no clouds are assumed to form outside the cooling radius.
This result of impact parameter as a function of $\vcir$
is similar to the observational result of Steidel et al. 
(see Steidel 1995). The observation shows
that the maximum impact parameter $D$ of their MgII systems
to the absorbing galaxies (having typically $L_K\ga 0.05 L_K^*$)
increases slowly with the K-band luminosity $L_K$ as 
$D\propto L_K^{0.15}$. Assuming the Tully-Fisher relation
$L_K\propto \vcir^4$, we have $D\propto \vcir^{0.6}$, which is
shown in the left panel of Fig.2 by the dotted line. We see  
that such a weak increase of $D$ with $\vcir$ 
can be accommodated in our model, arising from the value of the cooling
radius and the derived gas pressures in different halos.
The very small impact parameters in low $\vcir$ halos agrees with the fact
that MgII absorption systems are not 
commonly found in galaxies fainter than about $0.1L^*$ 
(Steidel 1995).

To see how different halos contribute to the MgII absorption,
we plot in the right panel of Fig.2 the cumulative number of MgII 
systems per unit redshift at $z=0.5$, produced in halos with $\vcir \le V_0$.
The figure shows that most MgII systems
are produced in halos with $\vcir=150$-$300\kms$.
The median $\vcir$ is about $200\kms$. 
The contribution to the total cross section made by
halos with $\vcir \la 150\kms$ is small, which again suggests
that MgII systems should not be commonly found in the halos of
dwarf galaxies.    

  The left panel in Fig.3 shows the total number of MgII systems and the
evolution with redshift, compared to the number of LLSs,
for two values of the metallicity. The relative number of MgII systems
to LLSs is correctly predicted when the metallicity is between
$Z= 0.1$-0.3, (compared with the observational results 
in Steidel \& Sargent 1992), and the evolution is also as observed 
if the metallicity in these halo clouds does not decrease very 
fast with redshift. For comparison, the total number of CIV systems 
in our model 
is also shown in the panel; the ratio to the number of MgII systems, 
and the evolution with redshift, is similar to what is observed.
Fig.3 (right panel) also compares the impact parameters
of CIV systems with those of MgII systems in different halos.
In small halos, $D$ is much larger
for CIV systems than for MgII systems. In fact,
the cross-section weighted average of $\vcir$ is only 
about $100\kms$ for the CIV systems, but as large as
about $200\kms$ for MgII systems, in the standard CDM model
considered here. Our model thus
predicts that, while MgII systems are mostly associated with
bright galaxies, many CIV systems should be found 
to be associated with small galaxies where $\nmgii$ is low. 
For the same reason, OVI systems should neither be commonly found
as photoionized clouds in halos with $\vcir\ga 150\kms$.
Some of such highly-ionized systems may arise from
collisionally ionized gas (e.g. in the hot phase)
or from an intermediate temperature phase
between the hot medium and the cold clouds.

\section {Discussion} 

  We have demonstrated that the simple halo model 
discussed in \S 2 can indeed reproduce the main properties of the  
observed galaxy-absorber connections. This is important,
because it means that the observed absorption systems 
can be understood in the current framework of galaxy formation.
  Much theoretical work remains to be done to see the plausibility
of our model and to make it more predictive, 
mainly to understand the physical processes determining the rate of
formation and destruction of gas clouds moving through the halo, the
distribution of cloud masses and velocities, etc. 
Further observations of absorption line systems and their relation to
galaxies should open a new era in our understanding of how galaxies
form, as the physical conditions of the gas in the halos around
galaxies, and the dependence of these conditions on galaxy luminosity
and morphology are unravelled.

\end{document}